\documentclass[%
 reprint,
superscriptaddress,
showpacs,
 amsmath,amssymb,
 aps,
]{revtex4-1}

\usepackage{rotating}
\usepackage{stackengine}
\usepackage{graphicx}
\usepackage{color}
\usepackage{verbatim}
\usepackage{subfigure}
\usepackage{tikz}
\usepackage{tkz-graph}

\definecolor{Blue}{rgb}{0,0,1}
\definecolor{Red}{rgb}{1,0,0}
\definecolor{Black}{rgb}{0,0,0}
\newcommand{\comments}[1]{{\color{Black} #1}}
\newcommand{\David}[1]{{\color{Black} #1}}
\newcommand{\Da}[1]{{\color{Black} #1}}
\newcommand{\B}[1]{{\color{Black} #1}}
\newcommand{\NewD}[1]{{\color{Black} #1}}

\begin{document}

\preprint{APS/123-QED}

\title{Scalable multimode entanglement based on efficient squeezing of propagation eigenmodes}

\author{David Barral} 
\email{david.barral@universite-paris-saclay.fr}
\affiliation{Centre de Nanosciences et de Nanotechnologies C2N, CNRS, Universit\'e Paris-Saclay, 10 boulevard Thomas Gobert, 91120 Palaiseau, France}
\author{Kamel Bencheikh}
\affiliation{Centre de Nanosciences et de Nanotechnologies C2N, CNRS, Universit\'e Paris-Saclay, 10 boulevard Thomas Gobert, 91120 Palaiseau, France}
\author{Juan Ariel Levenson}
\affiliation{Centre de Nanosciences et de Nanotechnologies C2N, CNRS, Universit\'e Paris-Saclay, 10 boulevard Thomas Gobert, 91120 Palaiseau, France}
\author{Nadia Belabas}
\affiliation{Centre de Nanosciences et de Nanotechnologies C2N, CNRS, Universit\'e Paris-Saclay, 10 boulevard Thomas Gobert, 91120 Palaiseau, France}

\begin{abstract}

\Da{Continuous-variable encoding of quantum information in the optical domain has recently yielded large temporal and spectral entangled states instrumental for quantum computing and quantum communication. We introduce a protocol for the generation of spatial multipartite entanglement based on phase-matching of a propagation eigenmode in a monolithic photonic device: the array of quadratic nonlinear waveguides. We theoretically demonstrate in the spontaneous parametric downconversion regime the generation of large multipartite entangled states useful for multimode quantum networks. Our protocol is remarkably simple and robust as it does not rely on specific values of coupling, nonlinearity or length of the sample.}
\end{abstract}

\date{November 18, 2020}

\maketitle 

\section{Introduction}

Optical networks play a key role in our everyday life as the substrate of a long-range communication grid: the internet. One goal of the blooming quantum technologies is the development of a quantum internet: an information web with unparalleled capabilities with respect to its current classical counterpart where information will be processed by quantum computers, transmitted in quantum secure channels, and routed towards quantum end nodes \cite{Wehner2018}. A must-have of the quantum internet is multipartite entanglement, where information is strongly correlated between the distributed nodes which compose the network \cite{Pirandola2015}. In addition, multipartite entanglement is the resource of a number of protocols in quantum communication, quantum sensing and quantum computing \cite{Hillery1999, vanLoock2001, Zhuang2018, Raussendorf2001}. Sources of multipartite entanglement are thus required in quantum networks and, particularly, light-based sources at telecom wavelengths are favored due to the current availability of large optical fiber networks. \Da{Such multipartite entanglement is elusive as it usually requires to create coherently and stabilize a large quantum objet prone to decoherence.}

Quantum information can be encoded in variables that take a continuous spectrum of eigenvalues --continuous variables (CV)-- \cite{Braunstein2005}. In the optical domain, the fluctuations of the field quadratures can be used as carriers of quantum information \cite{Andersen2010}. A number of table-top experiments have demonstrated CV quantum networks in the spatial, frequency and temporal domains \cite{Su2013, Armstrong2015, Chen2014, Cai2017, Yoshikawa2016, Takeda2019}. \Da{In the spatial domain the scaling up to usable systems} is far from feasible with bulk-optics systems. Scalability, stability and cost are issues that only well-established technologies like integrated and fiber optics can overcome \cite{Wang2019}. Generation of two-mode CV entanglement through bulk-integrated hybrid approaches has been explored \cite{Masada2015, Larsen2019} and, remarkably, a demonstration of a fully-on-chip source of CV bipartite entanglement has been recently proposed \cite{Lenzini2018}. \Da{Nevertheless, the mere transposition and extension of bulk-optics-based schemes --i.e. sequential squeezing and entanglement-- to larger number of modes is very demanding}. We present a simple and practical protocol for the generation of spatial multipartite entangled states of spontaneous parametric downconverted (SPDC) light \B{on chip} based on a currently-available technology: the array of $\chi^{(2)}$ nonlinear waveguides (ANW) \cite{Christodoulides2003, Iwanow2004, Setzpfandt2009, Solntsev2014}.
\begin{figure*}[t]
\centering
    \subfigure{\includegraphics[width=0.185\textwidth]{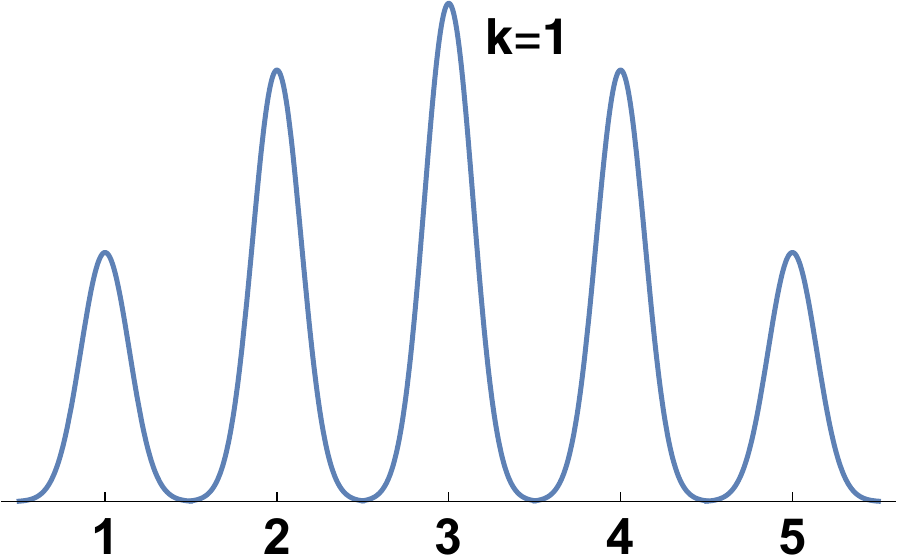}}
    \hspace {0cm}
    \subfigure{\includegraphics[width=0.185\textwidth]{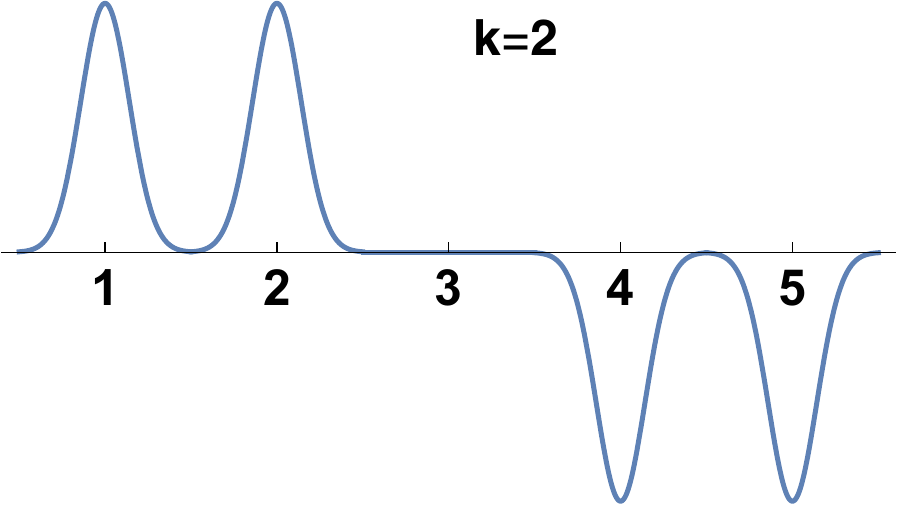}}
        \hspace {0cm}
    \subfigure{\includegraphics[width=0.185\textwidth]{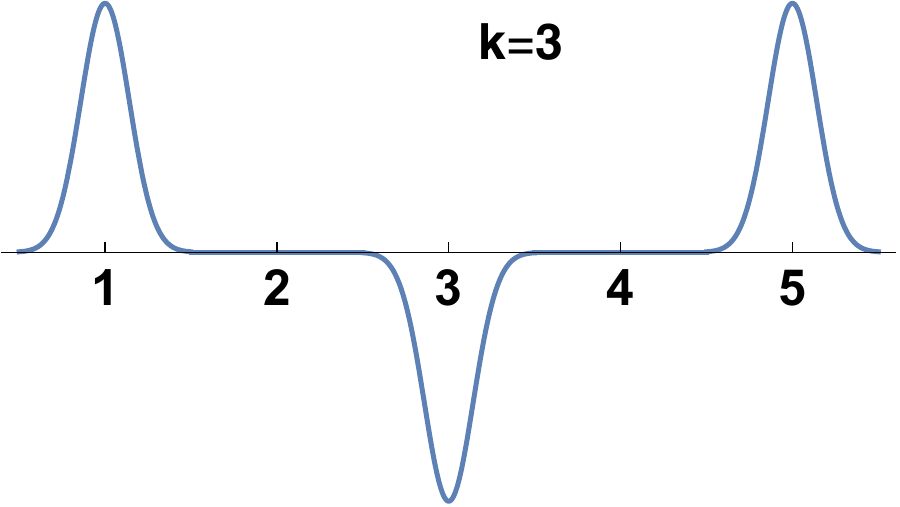}}
\hspace {0cm}\,
    \subfigure{\includegraphics[width=0.185\textwidth]{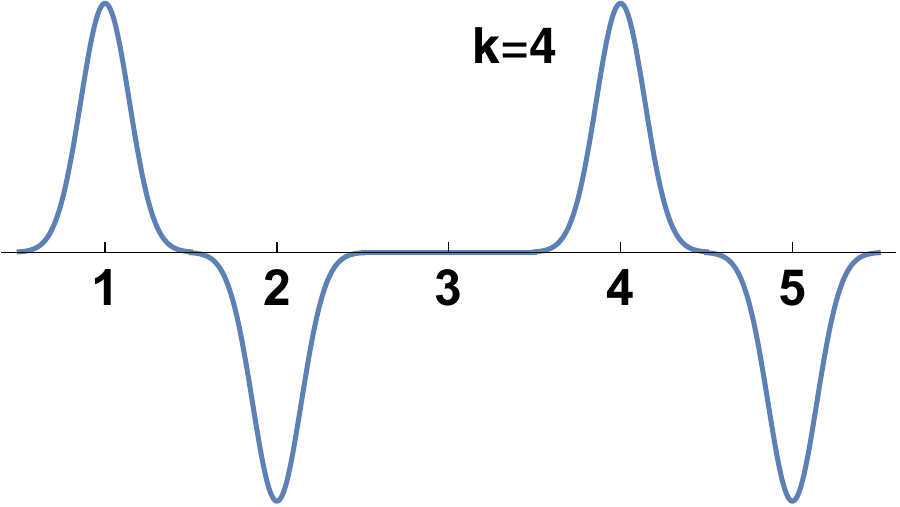}}
\hspace {0cm}\,
    \subfigure{\includegraphics[width=0.185\textwidth]{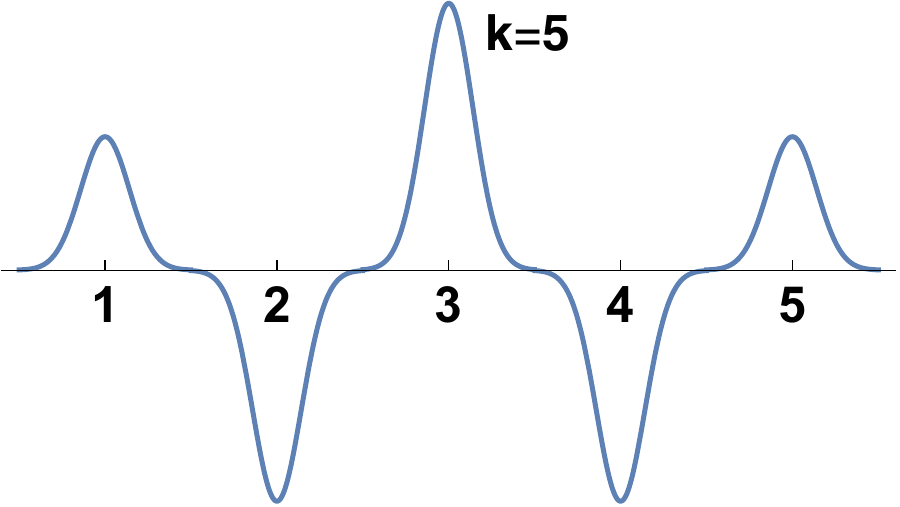}}
\vspace {0cm}\,
\hspace{0cm}\caption{\label{F1}\small{Sketch of the supermodes related to an array of linear waveguides with a homogeneous coupling profile $\vec{f}=\vec{1}$ and $N=5$. The horizontal axis stands for the individual modes. The propagation constants corresponding to each supermode are $\lambda=\{\sqrt{3}\, C_{0}, C_{0}, 0, -C_{0}, -\sqrt{3}\,C_{0}\}$. $k\equiv l=3$ is the zero supermode.}}
\end{figure*}

The distributed --i.e. simultaneous nonlinearity and evanescent coupling \cite{Barral2017}-- configuration of the ANW parallelizes multimode transformations by construction and hence embarks scalability. Our approach is \B{thus} conceptually different from the current strategy in spatially separated CV modes \cite{Masada2015, Larsen2019, Lenzini2018} and the state-of-the-art in path encoding \cite{Su2012, Armstrong2015, Chen2017, Wang2016, Wang2018}, where a N-mode interferometer is built to produce a specific quantum state through sequential two-mode transformations on a set of single or two-mode squeezed states. We showed that, in general, interferometer transformations can be approximated by suitable optimization of the ANW parameters \cite{Barral2020, Barral2020b}. Here we address the problem from a different perspective: we engineer quantum states harnessing collective modes of the system that are shape-invariant along propagation and have no equivalent in the standard sequential schemes. These propagation eigenmodes provide an analytical solution where the number of entangled modes naturally scales up with the number of waveguides. Using the properties of a particular collective mode that is phase-matched at all time, we build a large spatial entangled state with application to quantum networks. Physical ingredients such as propagation eigenmodes, analytical solutions building on symmetries, and phase matching thus give access to large and distributable states in a realistic system going beyond the technical limitations of bulk-optics inspired schemes to generate spatial entanglement. Our protocol is remarkably simple and robust as efficient building-up of this propagation eigenmode does not rely on specific values of coupling, nonlinearity or length of the sample. 

\NewD{The article is organized as follows: we first introduce the ANW and calculate the analytical solution for a specific pump profile in section II. We then use that solution to demonstrate and analyze multipartite entanglement among the individual modes which compose the phase-matched propagation eigenmode in section III. Finally, in section IV we discuss the feasibility and extensions of our protocol.}

\section{Dynamics of the array of nonlinear waveguides}

\Da{The ANW consists of $N$ identical $\chi^{(2)}$ waveguides. In each waveguide, an input harmonic field at frequency $\omega_{h}$ is downconverted into a signal field at frequency $\omega_{s}$. Pump-signal waves phase matching is produced only in the coupling region and is set to produce degenerate SPDC light. The generated signal fields are then coupled through evanescent tails in contrast to the pump fields which are not coupled due to a higher confinement in the waveguides. We deal with continuous variables $\hat{x}_{j}=(\hat{A}_{j}+\hat{A}_{j}^{\dag})$ and $\hat{y}_{j}=i (\hat{A}_{j}^{\dag}-\hat{A}_{j})$ which are, respectively, the orthogonal amplitude and phase quadratures corresponding to a monochromatic slowly-varying signal mode $\hat{A}_{j}\equiv \hat{A}_{j}(z, \omega_{s})$ propagating in the $j$th waveguide. \comments{The physical processes involved are described by a system of equations $d \hat{\xi}/{d z} = {\Delta}\, \hat{\xi}$ with $\hat{\xi}=(\hat{x}_{1}, \hat{y}_{1}, \dots, \hat{x}_{N}, \hat{y}_{N})^T$ and $z$ the coordinate along the direction of propagation \cite{Barral2017, Barral2020, Barral2020b}. ${\Delta}$ is a $2N\times2N$ block tridiagonal matrix made of $2\times 2$ nonlinear and evanescent-coupling blocks in the main and upper-lower diagonals, respectively.} These blocks involve $C_{j}=C_{0} f_{j}$, the linear coupling constant between nearest-neighbor modes $j$ and $j+1$, with $C_{0}$ the coupling strength and \Da{$\vec{f}=(f_{1}, \dots, f_{N-1})$} the coupling profile; and $\eta_{j}=g\, \alpha_{h,j}$, the effective nonlinear coupling constant corresponding to the $j$th waveguide, with $g$ the nonlinear constant proportional to $\chi^{(2)}$ and the spatial overlap of the signal and harmonic fields in each waveguide, and $\alpha_{h,j}$ a strong coherent pump field \cite{Barral2020b}. The nonlinear coupling constant $\eta_{j}$ can be tuned by means of a suitable set of pump phase and amplitude at each waveguide. In general, the propagation equations} can be solved numerically for a specific set of parameters $C_{j}$, $\eta_{j}$ and $N$, or even analytically if $N$ is small. Numerical or low-dimension analytical solutions do not provide much physical insight when increasing $N$. Remarkably, we have identified a case where analytical solutions are available for any $N$. This is the case of a flat pump profile, i.e. $\eta_{j}=\vert \eta \vert$. In this case the eigenmodes of the system are propagation eigenmodes --supermodes-- \cite{Kapon1984}. These eigenmodes form a basis and are represented by an orthogonal matrix \comments{$M\equiv M(\vec{f})$} with real elements $M_{k,j}$ \comments{\cite{Barral2020b}}. The individual and propagation supermode bases are related by $\hat{\xi}_{S,k}=\sum_{j=1}^{N} M_{k,j}\,\hat{\xi}_{j}$. The supermodes are orthonormal $\sum_{j=1}^{N}M_{k,j} M_{m,j}=\delta_{k,m}$, and the spectrum of eigenvalues is \comments{$\lambda_{k}\equiv \lambda_{k}(C_{0}, \vec{f})$}. \NewD{Figure \ref{F1} shows the supermodes of an array of 5 waveguides with a homogeneous coupling profile $\vec{f}=(f_{1}, \dots, f_{N})=\vec{1}$.} The solution of the propagation in this basis can be written as \Da{$\hat{\xi}_{S,k}(z)=S_{k}(z) \,\hat{\xi}_{S,k}(0)$, where}
\begin{equation}\label{S}
S_{k}(z)=\begin{pmatrix}
\cos(F_{k} z) & -e^{-r_{k}} \sin(F_{k} z) \\
 e^{r_{k}} \sin(F_{k} z) & \cos(F_{k} z)
\end{pmatrix},
\end{equation}
with $r_{k}=(1/2) \ln{[(\lambda_{k}+2\vert \eta \vert)/(\lambda_{k}-2\vert \eta \vert)]}$ and $F_{k}=\sqrt{\lambda_{k}^{2}-4\vert\eta\vert^{2}}$. For typical coupling strengths and pump powers found in quadratic ANW the condition $\vert \lambda_{k} \vert > 2 \vert \eta \vert$ is fulfilled. This regime is the relevant one for entanglement since, as the nonlinear interaction surpasses the linear coupling, the SPDC light tends to be more and more confined in the waveguide where \B{it} is created and then the ANW acts only as a group of individual squeezers \cite{Fiurasek2000}. We consider $F_{k} \in \mathbb{R}$ in the remainder of the article. Notably, the analytical solution of Equation (\ref{S}) is general for any evanescent coupling profile $\vec{f}$, any number of waveguides $N$ and any propagation distance $z$.

The quantum states generated in ANW are Gaussian. The most interesting observables are then the second-order moments of the quadrature operators, properly arranged in the covariance matrix ${V(z)}$ \cite{Adesso2014}. For a quantum state initially in vacuum, the elements of the covariance matrix $V(z)$ can be obtained from Equation (\ref{S}) using $V(z)=S(z) S^{T}(z)$ \Da{with $S(z)=$ diag$\{ S_{1}(z), \dots, S_{N}(z)\}$}. The covariance matrix elements in the supermode basis are then
\begin{align} \nonumber
V(x_{S,k}, x_{S,k})&= [\cosh{(r_{k})} + \sinh{(r_{k})} \cos{(2 F_{k} z)}] e^{-r_{k}},\\     \nonumber
V(y_{S,k}, y_{S,k})&= [\cosh{(r_{k})} - \sinh{(r_{k})} \cos{(2 F_{k} z)}] e^{+r_{k}}, \\   \label{Vs}
V(x_{S,k}, y_{S,k})&= \sinh{(r_{k})} \sin{(2 F_{k} z)}.
\end{align}
The squeezing ellipses for each supermode vary along propagation. Maximum (respectively null) squeezing are obtained periodically at distances $L_{k}=(2n-1)\pi/(2 F_{k})$ (respectively $L_{k}'=n \pi/ F_{k}$) that are different for each $k$th supermode, with \B{$n \in \mathbb{N}^{+}$}. The maximum value of squeezing available in the $k$th supermode is $e^{-2 r_{k}}=(\lambda_{k}-2\vert \eta \vert)/(\lambda_{k}+2\vert \eta \vert)$.

Instrumentally for our protocol, waveguide arrays with an odd number of identical waveguides present a supermode with zero eigenvalue --the zero supermode-- \cite{Efremidis2005}. This corresponds to the supermode with $k \equiv l=(N+1)/2$ and $\lambda_{l}=0$ (\NewD{see Figure \ref{F1}}) \cite{Barral2020b}. The elements of the covariance matrix for the zero supermode are thus
\begin{align}\nonumber
V(x_{S,l}, x_{S,l})&=V(y_{S,l}, y_{S,l})=\cosh{(4 \vert \eta \vert z)},\\ \label{VsL}
V(x_{S,l}, y_{S,l})&=\sinh{(4 \vert \eta \vert z)}. 
\end{align}
In contrast to the side supermodes $k\neq l$, the zero supermode noise efficiently builds up at all propagation distances and, notably, for large coupling strength the zero supermode is quickly dominant over the side supermodes \B{as} this is the only supermode which is phase-matched along propagation ($\lambda_{l}=0$). 

\B{In the diagonal supermode basis, no entanglement is available.} A simple change of basis takes Equation (\ref{Vs}) to the individual mode basis, corresponding to the individual waveguides output, obtaining
\begin{equation}
V(\xi_{i}, \xi_{j})=\sum_{k=1}^{N} M_{i,k} M_{j, k} V(\xi_{S,k}, \xi_{S,k}),  \label{Vz}
\end{equation}
with $\xi\equiv x$ or $y$.
Hence, the flat pump configuration generates quantum correlations --off-diagonal components of the covariance matrix-- between any pair $i$ and $j$ of individual modes, and thus full inseparability among individual modes can be produced. 
\begin{figure}[t]
  \centering
    {\includegraphics[width=0.44\textwidth]{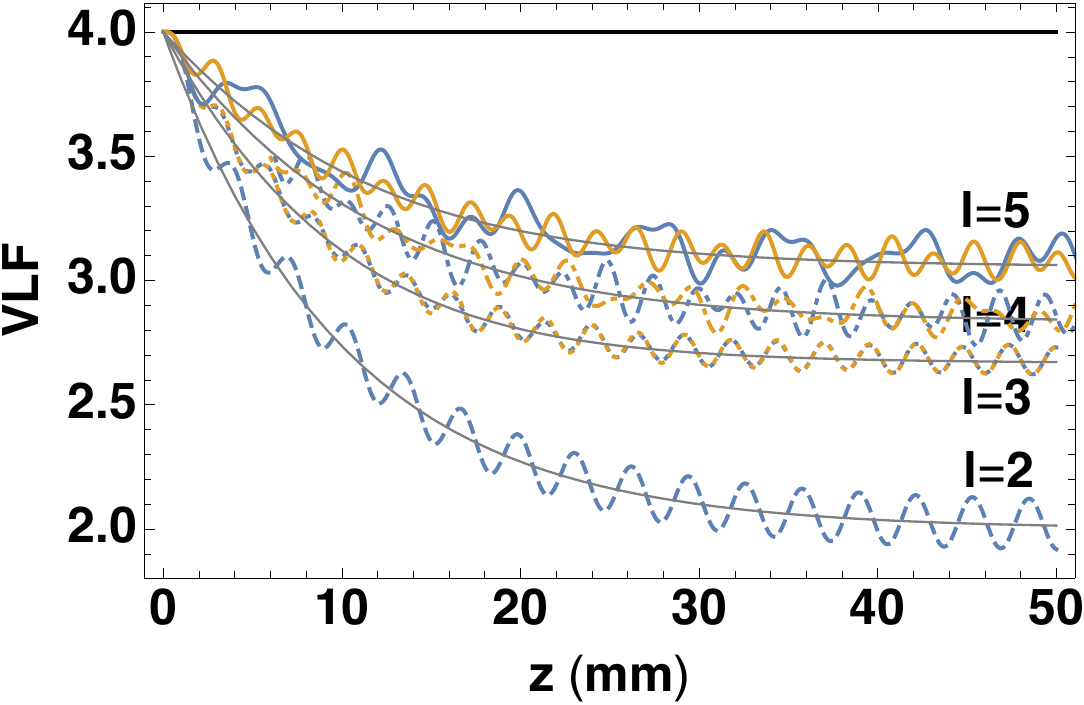}}
  \vspace {0cm}\,
\hspace{0cm}\caption{\label{F2}\small{Optimized van Loock - Furusawa inequalities $\text{VLF}(C_{0}, \vec{G}, z)$ for a small number of involved modes (color). Simultaneous values under the threshold value VLF=4 imply CV multipartite entanglement. Respective limits of large coupling $\text{VLF}(\infty, \vec{G}, z)$ are shown in solid gray. $C_{0}=0.70$ mm$^{-1}$. $\eta=0.025$ mm$^{-1}$.}}
\end{figure}

\section{Multipartite entanglement}

Measuring multipartite full inseparability in CV systems requires the simultaneous fulfillment of a set of conditions which leads to genuine multipartite entanglement when pure states are involved \cite{vanLoock2003}. This criterion, known as van Loock - Furusawa (VLF) inequalities, can be easily calculated from the elements of the covariance matrix $V$. Full $N$-partite inseparability is guaranteed if the following $N-1$ inequalities are simultaneously violated 
\small{ 
\begin{align}\nonumber
&\text{VLF}_{j}\equiv V[{x}_{j}({\theta_{j}}) - {x}_{j+1}(\theta_{j+1})] +  \\ \label{VLF}
&V[{y}_{j}(\theta_{j}) + {y}_{j+1}(\theta_{j+1})+\sum_{m\neq j,j+1}^{N} G_{m} \,{y}_{m}(\theta_{m})]  \geq 4, 
\end{align}}
where $\hat{x}_{j}({\theta_{j}})=\hat{x}_{j} \cos{(\theta_{j})}+\hat{y}_{j} \sin{(\theta_{j})}$ and $\hat{y}_{j}({\theta_{j}})=\hat{x}_{j}({\theta_{j}+\pi/2})$ are generalized quadratures \cite{Note1}. \B{$\vec{\theta}\equiv (\theta_{1}, \dots, \theta_{N})$ and $\vec{G}\equiv (G_{1}, \dots, G_{N})$ stand, respectively, for the LO phase and gain profiles  in multimode balanced homodyne detection (BHD) \B{that} can be set in order to minimize suitably the value of Equation (\ref{VLF}). We demonstrate below that the very efficient squeezing of the SPDC phase-matched supermode turns out to be a remarkably simple way of generating multipartite entanglement in ANW with an odd number of waveguides.}

Firstly, to gain insight about the multipartite entanglement generated with the flat pump configuration we tackle the limit of large coupling ($C_{0}\rightarrow \infty$). This limit is not physical as next-nearest-neighbor coupling should be in that case included in the model, but it gives us a clear insight on the dynamics of the system as the zero-supermode is then the dominant supermode generated in the array. In this limit an asymptotic lower bound on the violations of the VLF inequalities is obtained for the non-optimized case $\vec{G}=\vec{0}$ \David{\cite{Note2}}. The covariance matrix elements given by Equation (\ref{Vz}) for an array with odd number $N$ of waveguides in the limit of large coupling ($C_{0} \rightarrow \infty$) are significantly simplified to
\begin{align} \nonumber
&V(x_{i}, x_{j})=V(y_{i}, y_{j}) \rightarrow \delta_{i,j} + 2 M_{i,l} M_{j,l} \sinh^{2}(2 \vert \eta \vert z), \\ \label{LimitV}
&V(x_{i}, y_{j}) \rightarrow M_{i,l} M_{j,l} \sinh(4 \vert \eta \vert z).
\end{align}
Applying this result into the general expression for the VLF inequalities Equation (\ref{VLF}) without optimization ($G_{m\neq j, j+1}=0$) and using generalized quadratures with $\theta_{j}=0$ and $\theta_{j+1}=-\pi/2$ or $\pi/2$, we obtain
\begin{align}\nonumber
&\text{VLF}_{j}(\infty, \vec{0},z)=4-2(M_{j,l}^{2}+M_{j+1,l}^{2}) \\  \label{VLFj}
&\hspace{-0.2cm}+ (M_{j,l} \pm M_{j+1,l})^{2} \,e^{4 \vert \eta\vert z} + (M_{j,l} \mp M_{j+1,l})^{2} \,e^{-4 \vert \eta\vert z}\geq4,
\end{align}
where the upper (lower) signs corresponds to $\theta_{j+1}=-\pi/2 \,(\pi/2)$ and we have introduced the notation $\text{VLF} \equiv \text{VLF}(C_{0}, \vec{G}, z)$ for the sake of clarity. The best scenario in terms of violation of these inequalities corresponds to the case $M_{j,l}=\mp M_{j+1,l}$, for which we obtain
\begin{align}\nonumber
\text{VLF}_{j}&(\infty, \vec{0},z)=4-4M_{j,l}^{2}(1 - e^{-4 \vert \eta\vert z})<4 \quad \forall z > 0. 
\end{align}
\begin{figure}[t]
  \centering
    {\includegraphics[width=0.44\textwidth]{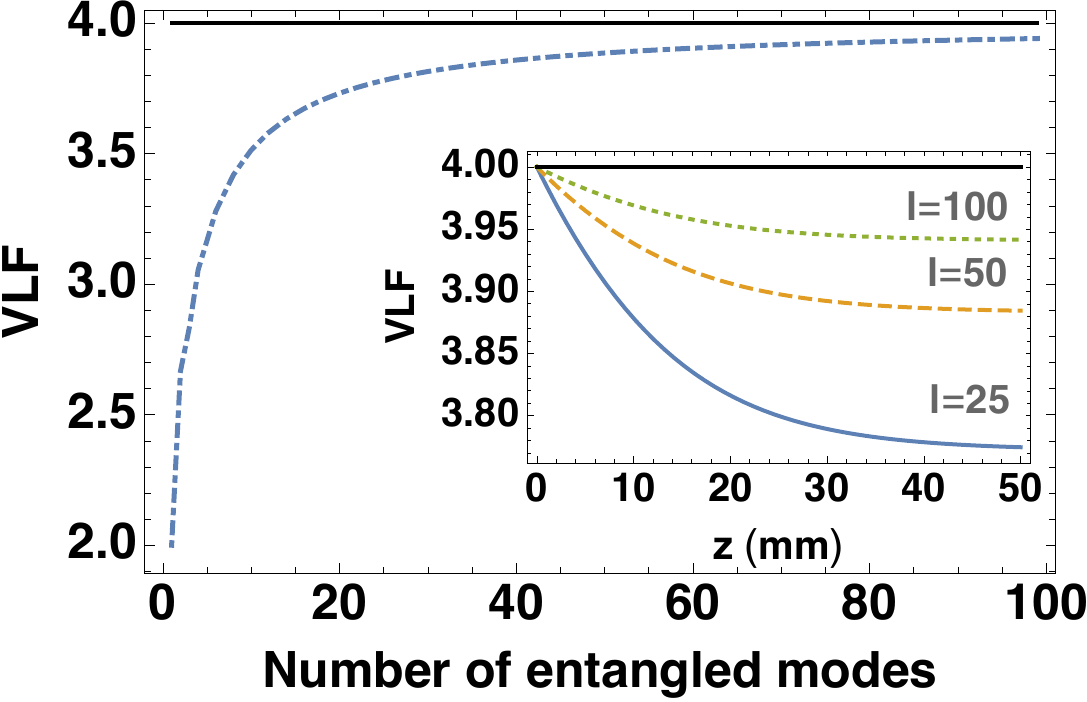}}
  \vspace {0cm}\,
\hspace{0cm}\caption{\label{F3}\small{\Da{Optimized van Loock - Furusawa inequalities in the limit of large coupling and propagation length $\text{VLF}(\infty, \vec{G}, \infty)$. Inset: value of the optimized inequalities for large number of involved modes in the limit of large coupling $\text{VLF}(\infty, \vec{G}, z)$. Simultaneous values under the threshold value VLF=4 imply CV multipartite entanglement. $\eta=0.025$ mm$^{-1}$.}}}
\end{figure}

Particularly, the coefficients of the zero supermode in an array with homogeneous coupling profile $\vec{f}=\vec{1}$ are given by $M_{j,l}=\sin(\frac{j\pi}{2})/\sqrt{l}$ \cite{Barral2019b}. Hence, mapping the mode $2j -1$ into the label $j$ of Equation (\ref{VLFj}), two solutions which maximize the violation of the separability conditions are obtained: \Da{a) the $l$ odd elements of the zero supermode satisfy $M_{2j-1,l}=-M_{2j+1,l}$ such that for a LO profile $(\theta_{2j-1},\theta_{2j+1})=(0,-\pi/2)$ multipartite entanglement is obtained among all the odd individual modes $\{1 ,3, 5, \dots N\}$, and b) the odd elements of the zero supermode satisfy $M_{2j-1,l}=M_{2j+3,l}$, thus for $(\theta_{2j-1},\theta_{2j+3})=(0, \pi/2)$ the multimode state is decoupled in two multipartite entangled states: $\{1, 5, 9, \dots\}$ and $\{3, 7, 11, \dots\}$}. Thus, the LO phase profile acts as an entanglement switch between two multimode entangled states. The individual modes propagating in the odd waveguides are fully inseparable in a measurement basis and separable in two parties --each fully inseparable-- in the other. Degenerate violation of the inseparability conditions is obtained in both cases
\begin{equation} \label{NOpti}
\text{VLF}(\infty, \vec{0}, z)=4\, \frac{(l-1)+e^{-4 \vert \eta\vert z}}{l} <4    \quad \forall \, j, l, z > 0.
\end{equation}
Hence, the strength of the violation of the VLF inequalities depends asymptotically only on the number of odd individual modes $l$ which make up the zero supermode. Moreover, the use of an optimized gain profile $\vec{G}\neq \vec{0}$ can only improve the above result. \B{We have indeed found an analytical optimized violation of the VLF conditions $\text{VLF}(\infty, \vec{G}, z)$} which depends also on the parity of $l$ (\NewD{see Appendix A}). \B{The correction produced optimizing $\vec{G}$ is negative and scales as $l^{-1}$ in the limit of a large number of modes.} Therefore, we have demonstrated that our protocol always produces multipartite entanglement in ANW in the limit of large coupling.

When finite coupling $C_{0}$ is taken into account, the side supermodes $k\neq l$ are present in the optimization of the VLF inequalities. This generates fluctuations around the value $\text{VLF}(\infty, \vec{G}, z)$. Figure \ref{F2} (color) shows one, two, three and four inequalities for arrays with, respectively, $l=$2, 3, 4 and 5 propagating modes obtained through minimization of Equations (\ref{VLF}) with a suitable gain profile $\vec{G}$ where we have used the analytical solutions of Equations (\ref{Vz}) \cite{Midgley2010}. The simultaneous violation of the inequalities  (VLF$_{j}<4$ in our notation) guarantees full inseparability, and since we deal with pure states the propagating signal modes are genuinely multipartite entangled. Interestingly, lower coupling strengths $C_{0} \rightarrow 0$ can lead to higher entanglement at specific lengths due to the increased strength of the side supermodes. The solid gray lines correspond to violations in the limit of large coupling $C_{0}\rightarrow \infty$ and optimized gain profiles $\vec{G}$ for each case. \B{Figure \ref{F2} thus demonstrates the validity of the asymptotic $\text{VLF}(\infty, \vec{G}, z)$ as a mean value of the real VLF inequalities that can be used to assess the possible entanglement generated in the array.}

Remarkably, quantum correlations are exhibited at any $z$ independently of the number $l$ of modes involved. \Da{Figure \ref{F3} depicts the relationship of the asymptotic value of VLF inequalities with the number of involved modes $l$ for $z\rightarrow \infty$. The evolution of multipartite entanglement along propagation for large number of modes ($l=$ 25, 50 and 100) is shown in the inset.} This figure demonstrates the scalability of our protocol. Noticeably, the asymptotic violation of the VLF inequalities in the double asymptotic limit ($C_{0}, z\rightarrow \infty$) given by $\text{VLF}(\infty, \vec{0}, \infty)=4 (l-1)/l$ is the same as that obtained in second harmonic generation (SHG) when a zero supermode is excited at the input of the ANW \cite{Barral2019b}. Unlike the SHG case where only the zero supermode is present, here the $k\neq l$ supermodes are involved in the production of entanglement. They increase the violation of the inequalities along $z$ through the use of optimized gains $\vec{G}$. The asymptotic behavior exhibited in Figures \ref{F1} and \ref{F2} appears as a consequence of tracing over the fields present in the even channels \cite{Barral2019b}.
\begin{figure}[t]
\begin{tikzpicture}
  \GraphInit[vstyle=Dijkstra]
  \SetVertexNormal[Shape=circle,FillColor=black!20]
  \Vertex[x=0,y=0,L=$1$]{A}    
  \Vertex[x=1.2,y=-1.2,L=$2$]{B}
  \Vertex[x=2.4,y=0,L=$3$]{C}  
  \Vertex[x=3.6,y=-1.2,L=$4$]{D}  
   \Vertex[x=4.8,y=0,L=$5$]{E}  
    \Vertex[x=6,y=-1.2,L=$6$]{F}  
  \tikzset{EdgeStyle/.style={-}}
  \Edge (A)(B)
   \Edge (B)(C)
   \Edge (C)(D)
   \Edge (A)(D)
   \Edge (D)(E)
    \Edge (B)(E)
    \Edge (E)(F)
     \Edge (A)(F)
      \Edge (C)(F)
\end{tikzpicture}
\caption{\label{F4}\small{\Da{A $l=6$ mode multiuser quantum channel (MQC) state obtained with the configuration $a)$. The nodes of the graph represent the modes of the MQC state, and the edges the EPR entanglement between pairs of modes.}}}
\end{figure}
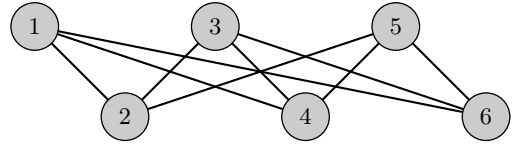

\Da{The multimode entangled states produced following this protocol have a special entanglement structure: they are multipartite entangled states with large bipartite entanglement between modes distributed in two groups. Using the configuration $a)$ in the limit of large coupling, the following two-mode EPR variances are the only nullifiers below their corresponding shot noise (2 in our notation) \cite{Koike2006}
\begin{align}\nonumber
&V(x_{i}-x_{j}')=V(y_{i}+y_{j}')=\frac{\text{VLF}(\infty, \vec{0}, z)}{2}<2, \forall \, i \,\text{odd}, \, j\,\text{even},   \\ \nonumber
&V(x_{i}'-x_{j})=V(y_{i}'+y_{j})=\frac{\text{VLF}(\infty, \vec{0}, z)}{2}<2, \forall  \, i \,\text{even}, \,  j\,\text{odd},
\end{align}
for all $z>0$, and where we have used Equation (\ref{NOpti}), labelled the odd modes $2j-1=1,3,\dots,N$ of the zero supermode as $j=1,2,\dots,l$ and used $x'=-y$, $y'=x$ for all even labels due to the chosen LO profile $\vec{\theta}=(0, -\pi/2, 0, -\pi/2, \dots)$. Thus, using Duan's inseparability criterion \cite{Duan2000} every odd $j$-mode is bipartite entangled with all the even $j$-modes and vice versa as shown in Figure \ref{F4} for $l=6$. This multipartite entangled state is a special class of multiuser quantum channel (MQC) state \cite{vanLoock2001}, the continuous-variable analogous to the qubit telecloning state \cite{Murao1999}. This state is particularly demanding for bulk-optics schemes as two multimode interferometers would be necessary \cite{vanLoock2001}. The MQC can be symmetric if the number of nodes available $l$ is even or asymmetric if is odd. Every node of the upper (lower) side can be used to teleclone quantum information to all the nodes of the other side. A similar result is obtained with the configuration $b)$ (\NewD{see Appendix B}). Thus, our scheme can be used as a resource for multiuser unassisted quantum networks.}

The above analysis based on two-mode entanglement can be in principle improved further by elucidating the canonical graph of our state. Using the theory developed in \cite{Menicucci2011} we have found however that the graph of our multipartite entangled state is not converging to the graph of a CV cluster state in the limit of large squeezing (\NewD{see Appendix C}).

\section{Feasibility and outlook}

\Da{\B{Finally, our protocol yields significant and useful entanglement over a wide range of number of modes. This scheme is very appealing for the generation of scalable multipartite entanglement since it relies on coupling $C_{0}$ and nonlinearity $g$ within the array, but not on specific values of these parameters. Feasibility of entanglement generation in ANW has been discussed in detail in \cite{Barral2020}. Noticeably, a squeezing level as high as -6.3 dB has been recently demonstrated in a PPLN waveguide in the cw regime \cite{Kashiwazaki2020}. Propagation losses have a small impact on squeezing and entanglement assuming typical values in PPLN waveguides and sample lengths \cite{Lenzini2018, Mondain2019}.} Furthermore, our method gives insight on further extension of the possibilities of the ANW for a resource-efficient generation of large entangled states instrumental for more advanced communication and computing protocols. We point at several enablers: i) our
scheme is platform independent and can thus benefit of highly efficient nonlinear waveguides \cite{Lenzini2018,Kashiwazaki2020, Mondain2019}, ii) phasematching every supermode of the array using supermode quasi-phasematching to optimize entanglement \cite{Barral2019c}, iii) large GHZ and cluster states can be obtained via further optimization of pump, coupling and LO profiles \cite{Barral2020, Barral2020b}, iv) our scheme is compatible with a pulsed pump, hence adapted to temporal multiplexing \cite{Larsen2019b, Asavanant2019}, and v) non-Gaussian quantum advantage can be obtained by means of weakly coupled waveguides, in such a way that the detection of a single photon in any of these waveguides de-Gaussifies the multimode entangled state \cite{Ourjoumtsev2009, Ra2020}. }

\section*{Acknowledgements} We thank N. C. Menicucci for insight on graphical calculus. This work was supported by the Agence Nationale de la Recherche through the INQCA project (Grants No. PN-II-ID-JRP-RO-FR-2014-0013 and No. ANR-14-CE26-0038), the Paris Ile-de-France region in the framework of DIM SIRTEQ through the project ENCORE, and the Investissements d'Avenir program (Labex NanoSaclay, reference ANR-10-LABX-0035).

\appendix

\section{Derivation of $\text{VLF}_{j}(\infty, \vec{G}, z)$ shown in Figures $\ref{F2}$ and $\ref{F3}$}
\Da{In the following we show how $\text{VLF}_{j}(\infty, \vec{G}, z)$ is obtained. We start writing the VLF inequalities of Equation (\ref{VLF}) in the following way
\begin{align}\nonumber
\text{VLF}_{j}\equiv &V[{x}_{j}({\theta_{j}}) - {x}_{j+1}(\theta_{j+1})] +V[{y}_{j}(\theta_{j}) + {y}_{j+1}(\theta_{j+1}) \\  \nonumber
&+\sum_{o\neq j,j+1}^{O} G_{o} \,{y}_{o}(\theta_{o})+\sum_{e\neq j,j+1}^{E} G_{e} \,{y}_{e}(\theta_{e})] \geq 4,
\end{align}}
where we have chosen generalized phase quadratures for the odd ($o$) auxiliary modes and generalized amplitude quadratures for the even ($e$) auxiliary modes, and $O$ and $E$ are the total number of odd and even auxiliary modes ($O+E=N-2$), respectively. Taking into account only the $l$ odd elements of the zero supermode  in the limit of infinite coupling, this equation gives 
\begin{align}\nonumber
&\text{VLF}_{j}(\infty, \vec{G}, z) = \\ \nonumber
&\text{VLF}(\infty, \vec{0}, z)+ \frac{e^{4 \vert \eta \vert z} }{2 l}(\sum_{o\neq j,j+1}^{O} G_{o} - \sum_{e\neq j,j+1}^{E} G_{e})^{2}+\\ \nonumber
&\frac{e^{-4 \vert \eta \vert z} }{2 l}[(\sum_{o\neq j,j+1}^{O} G_{o} + \sum_{e\neq j,j+1}^{E} G_{e})^{2} + 4(\sum_{o\neq j,j+1}^{O} G_{o} + \sum_{e\neq j,j+1}^{E} G_{e})],
\end{align}
where we have used Equation (\ref{VLFj}) of the main text and the mapping: mode $2j-1 \rightarrow$ label $j$. The VLF inequalities are minimized if $\sum_{o\neq j,j+1}^{O} G_{o} = \sum_{e\neq j,j+1}^{E} G_{e}$. The symmetries of the system --flat pump profile, homogeneous coupling profile-- suggest that all inequalities should be degenerate. To that end, we set equally all the odd (even) gains such that $\sum_{o\neq j,j+1}^{O} G_{o}=O \,G_{o}$ ($\sum_{e\neq j,j+1}^{E} G_{e}=E \,G_{e}$). Our numerical simulations support this assumption. The optimization condition is thus $G_{e}=(O/E) \,G_{o}$ and the VLF inequalities read
\begin{align}\nonumber
\text{VLF}_{j}(\infty, \vec{G}, z) = &\text{VLF}(\infty, \vec{0}, z)- \frac{4 O}{l} (1-e^{-4 \vert \eta \vert z}) G_{o} \\ \nonumber
&+O[\frac{O+E}{E}-\frac{2 O}{l} (1-e^{-4 \vert \eta \vert z})]G_{o}^{2}.
\end{align}
The parameter $G_{o}$ which minimizes this equation is obtained by solving $\text{d} \text{VLF}_{j}(\infty, \vec{G}, z) / \text{d} G_{o}=0$ for $G_{o}$. The VLF inequalities  using this optimized gain are hence
\begin{align}\nonumber
&\text{VLF}_{j}(\infty, \vec{G}, z) = \\ \nonumber
&\text{VLF}(\infty, \vec{0}, z) - \frac{4 \,O E}{l} \frac{(1-e^{-4 \vert \eta \vert z})^{2}}{l(O+E)-2\, O E (1-e^{-4 \vert \eta \vert z})}.
\end{align}
The VLF inequalities depend thus on the parity of $l$ --the total number of involved modes--. Note that the total number of auxiliary modes is $O+E=l-2$. If $l$ is odd, $E=(l-3)/2$ and $O=(l-1)/2$, and we obtain
\begin{align}\nonumber
&\text{VLF}(\infty, \vec{G}, z) = \text{VLF}(\infty, \vec{0}, z) \\ \nonumber
&- \frac{2(l^{2}-4l+3)}{l} \frac{(1-e^{-4 \vert \eta \vert z})^{2}}{2l(l-2)-(l^{2}-4l+3)(1-e^{-4 \vert \eta \vert z}) }<4 \\ \nonumber
&\hspace{6.6cm} \forall \, j, z , l \, \text{odd}.
\end{align}
Note that there is optimization only for $l > 3$. If $l$ is even, $E=O=(l-2)/2$, and we obtain
\begin{align}\nonumber
&\text{VLF}(\infty, \vec{G}, z) = \\ \nonumber
&\text{VLF}(\infty, \vec{0}, z) - \frac{2(l-2) }{l} \frac{(1-e^{-4 \vert \eta \vert z})^{2}}{2l-(l-2)(1-e^{-4 \vert \eta \vert z}) }<4 \\ \nonumber
&\hspace{5.85cm} \forall \,j, z , l \, \text{even}.
\end{align}

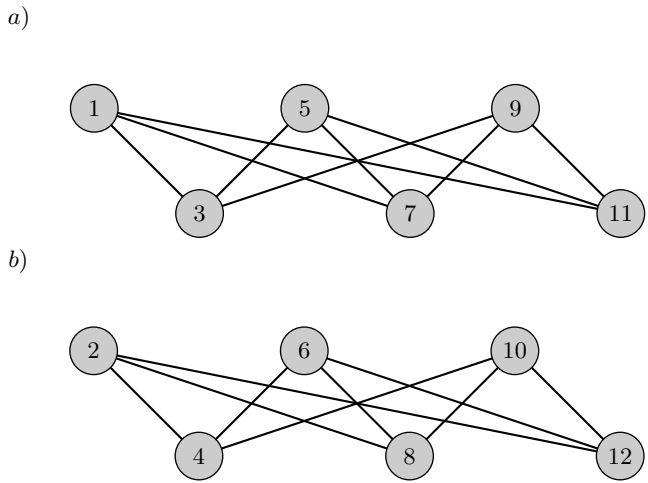
\begin{figure}[h]
\begin{center}
\begin{tikzpicture}
\node at (-1,1.2) {$a)$};
  \GraphInit[vstyle=Dijkstra]
  \SetVertexNormal[Shape=circle,FillColor=black!20]
  \Vertex[x=0,y=0,L=$1$]{A}    
  \Vertex[x=1.4,y=-1.4,L=$3$]{B}
  \Vertex[x=2.8,y=0,L=$5$]{C}  
  \Vertex[x=4.2,y=-1.4,L=$7$]{D}  
   \Vertex[x=5.6,y=0,L=$9$]{E}  
    \Vertex[x=7,y=-1.4,L=$11$]{F}  
  \tikzset{EdgeStyle/.style={-}}
  \Edge (A)(B)
   \Edge (B)(C)
   \Edge (C)(D)
   \Edge (A)(D)
   \Edge (D)(E)
    \Edge (B)(E)
    \Edge (E)(F)
     \Edge (A)(F)
      \Edge (C)(F)
\end{tikzpicture} \\
\begin{tikzpicture}
\node at (-1,1.2) {$b)$};
  \GraphInit[vstyle=Dijkstra]
  \SetVertexNormal[Shape=circle,FillColor=black!20]
  \Vertex[x=0,y=0,L=$2$]{A}    
  \Vertex[x=1.4,y=-1.4,L=$4$]{B}
  \Vertex[x=2.8,y=0,L=$6$]{C}  
  \Vertex[x=4.2,y=-1.4,L=$8$]{D}  
   \Vertex[x=5.6,y=0,L=$10$]{E}  
    \Vertex[x=7,y=-1.4,L=$12$]{F}  
  \tikzset{EdgeStyle/.style={-}}
  \Edge (A)(B)
   \Edge (B)(C)
   \Edge (C)(D)
   \Edge (A)(D)
   \Edge (D)(E)
    \Edge (B)(E)
    \Edge (E)(F)
     \Edge (A)(F)
      \Edge (C)(F)
\end{tikzpicture}
\caption{\label{F5}\small{\Da{"Odd" (a) and "even" (b) 6-mode MQC states obtained with the configuration b). The nodes of the graph represent the modes of the MQC state, and the edges the EPR entanglement between pairs of modes.}}}
\end{center}
\end{figure}

\section{Entanglement structure of the generated quantum state for condition b}

We disclose here the entanglement structure of the generated quantum states corresponding to the second solution which maximize the violation of the separability conditions. We tackle the limit of large coupling.

{\bf Condition b:} The $l$ odd elements of the zero supermode satisfy $M_{2j-1,l}=M_{2j+3,l}$. The multimode state is decoupled in two multipartite entangled states: $\{1, 5, 9, 13, \dots\}_{o}$ and $\{3, 7, 11, 15, \dots\}_{e}$ for two degenerate LO phase profiles $\vec{\theta}_{o,e}=(0, \pi/2, 0, \pi/2, \dots)$.

We start labelling the odd modes $2j-1=1,3,\dots,N$ of the zero supermode as $j=1,2,\dots,l$. The decoupling with this labelling is given in terms of odd $j=1,3,5, \dots$ and even $j=2,4,6, \dots$. The following two-mode EPR variances are the only nullifiers below their corresponding shot noise (2 in our notation) \cite{Koike2006}
\begin{align}\nonumber
V(x_{i}-x_{j})=&V(y_{i}+y_{j})=2\frac{(l-1)+e^{-4 \vert \eta\vert z}}{l}<2, \\ \nonumber
&\forall \, z >0, \,i=\{1,5,9,\dots\},\,j=\{3,7,11,\dots\}, \\  \nonumber
 V(x_{i}'-x_{j}')=&V(y_{i}'+y_{j}')=2\frac{(l-1)+e^{-4 \vert \eta\vert z}}{l}<2, \\ \nonumber
 &\forall \, z >0, \,i=\{2,6,10,\dots\},\,j=\{4,8,12,\dots\},
\end{align}
where $x'=y, y'=-x$ for all even labels due to the chosen LO profile. Thus, using Duan's inseparability criterion \cite{Duan2000} we have two sets of modes: the odd set where each mode $j=1,5, 9,\dots$ is bipartite entangled with the modes $j=3, 7, 11, \dots$ and vice versa (Figure \ref{F5}a); and the even set where each mode $j=2,6, 10,\dots$ is bipartite entangled with the modes $j=4, 8, 12, \dots$ and vice versa (Figure \ref{F5}b). We have now two decoupled multiuser quantum channel (MQC) states \cite{vanLoock2001}. Each set is genuinely multipartite entangled, as demonstrated in the main text.

\section{Canonical graph}
From the covariance matrix elements in the limit of large coupling given by Equation (6), we build the complex-weighted adjacency matrix ${\bf Z}={\bf V}+i \,{\bf U}$ \cite{Menicucci2011}, where
\begin{align} \nonumber
{\bf V}=&\frac{1}{l} \tanh(4\vert \eta \vert z) 
\begin{pmatrix}
1 & -1 & 1 & \dots \\
-1 & 1 & -1& \dots \\
1 & -1 & 1 & \dots \\
\vdots & \vdots & \vdots
\end{pmatrix}, \\ \nonumber
\quad {\bf U}=&\frac{1}{l} \text{sech}(4\vert \eta \vert z) 
\begin{pmatrix}
1 & -1 & 1 & \dots \\
-1 & 1 & -1& \dots \\
1 & -1 & 1 & \dots \\
\vdots & \vdots & \vdots
\end{pmatrix} \\ \nonumber
&\hspace{0.8cm}+\frac{1}{l} 
\begin{pmatrix}
l-1 & 1 & -1 & \dots \\
1 & l-1 & 1& \dots \\
-1 & 1 & l-1 & \dots \\
\vdots & \vdots & \vdots
\end{pmatrix}.
\end{align}
In the limit of large squeezing $4 \vert \eta \vert z\rightarrow \infty$, ${\bf V}$ is given by
\begin{equation} \nonumber
{\bf V}\rightarrow \frac{1}{l}
\begin{pmatrix}
1 & -1 & 1 & \dots \\
-1 & 1 & -1& \dots \\
1 & -1 & 1 & \dots \\
\vdots & \vdots & \vdots
\end{pmatrix}\equiv {\bf V}_{\infty}.
\end{equation}
${\bf V}_{\infty}$ is the canonical graph of our state approximated by ${\bf Z}$ if the error in the approximation $\text{tr}{\,\bf U}$ vanishes in the limit of large squeezing. However, the error in the approximation in this limit is equal to the number of unused even modes times the shot noise: $\text{tr}{\,\bf U}\rightarrow l-1 \geq 1$. We checked that this large value of the error in the limit of large squeezing remained even when incorporating local phase shifts \cite{Menicucci2011}. Thus the state can not be considered a cluster state close to the canonical graph ${\bf V}_{\infty}$.

\section*{}

\end{document}